\documentclass[preprint]{aastex62}
\usepackage{amsbsy}

\newcommand{\affOSU}{Department of Astronomy, Ohio State University, 140 W. 18th Ave., Columbus, OH 43210, USA}
\newcommand{\affOAUW}{Astronomical Observatory, University of Warsaw, Al. Ujazdowskie 4, 00-478 Warszawa, Poland}
\newcommand{\affWARWICK}{Department of Physics, University of Warwick, Coventry CV4 7AL, UK}
\newcommand{\affTAS}{School of Natural Sciences, University of Tasmania, Private Bag 37 Hobart, Tasmania 7001 Australia}
\newcommand{\affSORB}{Sorbonne Universit\'e, UPMC Universit\'e Paris 6 et CNRS, UMR 7095, Institut d'Astrophysique de Paris, 98 bis Bd Arago, F-75014 Paris, France}
\newcommand{\affBORD}{Laboratoire d'Astrophysique de Bordeaux, Univ. Bordeaux, CNRS, B18N, all\'ee Geoffroy Saint-Hilaire, F-33615 Pessac, France}
\newcommand{\affKAVLI}{Kavli Institute for Astronomy and Astrophysics, Peking University, Yi He Yuan Road 5, Hai Dian District, Beijing 100871, China}
\newcommand{\affCHUNGBUK}{Department of Physics, Chungbuk National University, Cheongju 28644, Republic of Korea}
\newcommand{\affARI}{Steward Observatory, University of Arizona, Tucson, AZ 85721, USA}
\newcommand{\affHARV}{Harvard-Smithsonian Center for Astrophysics, 60 Garden Street, Cambridge, MA 02138, USA}
\newcommand{\affKASI}{Korea Astronomy and Space Science Institute, 776 Daedukdae-ro, Yuseong-gu, Daejeon 34055, Korea}
\newcommand{\affMAX}{Max-Planck-Institute for Astronomy, K\"{o}igstuhl 17, 69117 Heidelberg, Germany}

\newcommand{\E}{{\rm E}}

\received{ABCDEF XX, 2019}
\revised{ABCDEF XX, 2019}
\accepted{ABCDEF XX, 2019}
\submitjournal{ApJ}

\shorttitle{OGLE-2012-BLG-0838Lb}
\shortauthors{Poleski et al.}

\begin{document}

\title{A Wide-orbit Exoplanet OGLE-2012-BLG-0838Lb}

\correspondingauthor{Rados\l{}aw Poleski}
\email{rpoleski@astrouw.edu.pl}

\author[0000-0002-9245-6368]{R.~Poleski}
\affiliation{\affOSU}\affiliation{\affOAUW}
\author[0000-0002-5843-9433]{Daisuke Suzuki}
\affil{Institute of Space and Astronautical Science, Japan Aerospace Exploration Agency, 3-1-1 Yoshinodai, Chuo, Sagamihara, Kanagawa 252-5210, Japan}
\author[0000-0001-5207-5619]{A.~Udalski}
\affiliation{\affOAUW}
\author{Xiaojia~Xie}
\affiliation{\affKAVLI} 
\author[0000-0001-9481-7123]{J.~C.~Yee}
\affiliation{\affHARV}
\author[0000-0003-2302-9562]{Naoki Koshimoto}
\affiliation{Department of Astronomy, Graduate School of Science, The University of Tokyo, 7-3-1 Hongo, Bunkyo-ku, Tokyo 113-0033, Japan}
\affiliation{National Astronomical Observatory of Japan, 2-21-1 Osawa, Mitaka, Tokyo 181-8588, Japan}
\affiliation{Astrophysics Science Division, NASA Goddard Space Flight Center, Greenbelt, MD 20771, USA}
\author[0000-0003-0395-9869]{B.~S.~Gaudi}
\affiliation{\affOSU}
\author{A.~Gould}
\affiliation{\affMAX}\affiliation{\affKASI}\affiliation{\affOSU}
\collaboration{(leading authors)}

\author[0000-0002-2335-1730]{J.~Skowron}
\affiliation{\affOAUW}
\author[0000-0002-0548-8995]{M.~K.~Szyma\'nski}
\affiliation{\affOAUW}
\author[0000-0002-7777-0842]{I.~Soszy\'nski}
\affiliation{\affOAUW} 
\author[0000-0002-2339-5899]{P.~Pietrukowicz}
\affiliation{\affOAUW}
\author[0000-0003-4084-880X]{S.~Koz\l{}owski}
\affiliation{\affOAUW}
\author[0000-0002-9658-6151]{\L{}.~Wyrzykowski}
\affiliation{\affOAUW}
\author[0000-0001-6364-408X]{K.~Ulaczyk}
\affiliation{\affOAUW}\affiliation{\affWARWICK}
\collaboration{(OGLE collaboration)}

\author{Fumio Abe}
\affiliation{Institute for Space-Earth Environmental Research, Nagoya University, Furo-cho, Chikusa, Nagoya, Aichi 464-8601, Japan}
\author{Richard K. Barry}
\affiliation{Astrophysics Science Division, NASA Goddard Space Flight Center, Greenbelt, MD 20771, USA}
\author[0000-0001-8043-8413]{David. P. Bennett}
\affiliation{Astrophysics Science Division, NASA Goddard Space Flight Center, Greenbelt, MD 20771, USA}
\affiliation{Department of Astronomy, University of Maryland, College Park, MD 20742, USA}
\author{Aparna Bhattacharya}
\affiliation{Astrophysics Science Division, NASA Goddard Space Flight Center, Greenbelt, MD 20771, USA}
\affiliation{Department of Astronomy, University of Maryland, College Park, MD 20742, USA}
\author{Ian A. Bond}
\affiliation{Institute of Information and Mathematical Sciences, Massey university, Private Bag 102-904, North Shore Mail Centre, Auckland, New Zealand}
\author{Martin Donachie}
\affiliation{Department of Physics, University of Auckland, Private Bag 92019, Auckland, New Zealand}
\author{Hirosane Fujii}
\affiliation{Institute for Space-Earth Environmental Research, Nagoya University, Furo-cho, Chikusa, Nagoya, Aichi 464-8601, Japan}
\author[0000-0002-4909-5763]{Akihiko Fukui}
\affiliation{Department of Earth and Planetary Science, Graduate School of Science, The University of Tokyo, 7-3-1 Hongo, Bunkyo-ku, Tokyo 113-0033, Japan}
\affiliation{Instituto de Astrof\'isica de Canarias, V\'ia L\'actea s/n, E-38205 La Laguna, Tenerife, Spain}
\author{Yoshitaka Itow}
\affiliation{Institute for Space-Earth Environmental Research, Nagoya University, Furo-cho, Chikusa, Nagoya, Aichi 464-8601, Japan}
\author{Yuki Hirao}
\affiliation{Department of Earth and Space Science, Graduate School of Science, Osaka University, 1-1 Machikaneyama, Toyonaka, Osaka 560-0043, Japan}
\affiliation{Astrophysics Science Division, NASA Goddard Space Flight Center, Greenbelt, MD 20771, USA}
\author{Yuhei Kamei}
\affiliation{Institute for Space-Earth Environmental Research, Nagoya University, Furo-cho, Chikusa, Nagoya, Aichi 464-8601, Japan}
\author{Iona Kondo}
\affiliation{Department of Earth and Space Science, Graduate School of Science, Osaka University, 1-1 Machikaneyama, Toyonaka, Osaka 560-0043, Japan}
\author{Man Cheung Alex Li}
\affiliation{Department of Physics, University of Auckland, Private Bag 92019, Auckland, New Zealand}
\author{Yutaka Matsubara}
\affiliation{Institute for Space-Earth Environmental Research, Nagoya University, Furo-cho, Chikusa, Nagoya, Aichi 464-8601, Japan}
\author[0000-0001-9818-1513]{Shota Miyazaki}
\affiliation{Department of Earth and Space Science, Graduate School of Science, Osaka University, 1-1 Machikaneyama, Toyonaka, Osaka 560-0043, Japan}
\author{Yasushi Muraki}
\affiliation{Institute for Space-Earth Environmental Research, Nagoya University, Furo-cho, Chikusa, Nagoya, Aichi 464-8601, Japan}
\author{Masayuki Nagakane}
\affiliation{Department of Earth and Space Science, Graduate School of Science, Osaka University, 1-1 Machikaneyama, Toyonaka, Osaka 560-0043, Japan}
\author[0000-0003-2388-4534]{Cl\'ement Ranc}
\affiliation{Astrophysics Science Division, NASA Goddard Space Flight Center, Greenbelt, MD 20771, USA}
\author[0000-0001-5069-319X]{Nicholas J. Rattenbury}
\affiliation{Department of Physics, University of Auckland, Private Bag 92019, Auckland, New Zealand}
\author{Yuki K. Satoh}
\affiliation{Department of Earth and Space Science, Graduate School of Science, Osaka University, 1-1 Machikaneyama, Toyonaka, Osaka 560-0043, Japan}
\author{Hikaru Shoji}
\affiliation{Department of Earth and Space Science, Graduate School of Science, Osaka University, 1-1 Machikaneyama, Toyonaka, Osaka 560-0043, Japan}
\author{Haruno Suematsu}
\affiliation{Department of Earth and Space Science, Graduate School of Science, Osaka University, 1-1 Machikaneyama, Toyonaka, Osaka 560-0043, Japan}
\author{Denis J. Sullivan}
\affiliation{School of Chemical and Physical Sciences, Victoria University, Wellington, New Zealand}
\author{Takahiro Sumi}
\affiliation{Department of Earth and Space Science, Graduate School of Science, Osaka University, 1-1 Machikaneyama, Toyonaka, Osaka 560-0043, Japan}
\author{Paul J. Tristram}
\affiliation{University of Canterbury Mt. John Observatory, P.O. Box 56, Lake Tekapo 8770, New Zealand}
\author{Takeharu Yamakawa}
\affiliation{Institute for Space-Earth Environmental Research, Nagoya University, Furo-cho, Chikusa, Nagoya, Aichi 464-8601, Japan}
\author{Tsubasa Yamawaki}
\affiliation{Department of Earth and Space Science, Graduate School of Science, Osaka University, 1-1 Machikaneyama, Toyonaka, Osaka 560-0043, Japan}
\author{Atsunori Yonehara}
\affiliation{Department of Physics, Faculty of Science, Kyoto Sangyo University, Kyoto 603-8555, Japan}
\collaboration{(MOA collaboration)}

\author[0000-0002-2641-9964]{C.~Han}
\affiliation{\affCHUNGBUK} 
\author[0000-0002-1027-0990]{Subo~Dong}
\affiliation{\affKAVLI}
\author[0000-0002-1384-0063]{K.~M.~Morzinski}
\affiliation{\affARI} 
\author{J.~R.~Males}
\affiliation{\affARI}
\author[0000-0002-2167-8246]{L.~M.~Close}
\affiliation{\affARI} 
\author[0000-0003-1435-3053]{R.~W.~Pogge}
\affiliation{\affOSU}
\author[0000-0003-0014-3354]{J.-P.~Beaulieu}
\affiliation{\affTAS}\affiliation{\affSORB}
\author{J.-B.~Marquette}
\affiliation{\affBORD}\affiliation{\affSORB}
\nocollaboration

\begin{abstract}
We present the discovery of a planet on a very wide orbit in 
the microlensing event OGLE-2012-BLG-0838.  The signal of the planet is well 
separated from the main peak of the event and the planet-star 
projected separation is found to be twice the Einstein ring radius, 
which corresponds to a projected separation of $\approx 4~{\rm AU}$.
Similar planets around low-mass stars are very hard to find using 
any technique other than microlensing. 
We discuss microlensing model fitting in detail and 
discuss the prospects for measuring the mass and distance of the lens system directly. 
\end{abstract}


\section{Introduction} \label{sec:intro}

The exoplanets known today show a large degree of diversity. 
For example, we now know 
a planetary system orbiting a pulsar \citep[PSR1257+12;][]{wolszczan92},
extremely short-period planets \citep[55 Cnc e;][]{winn11},
planets with extremely high surface temperatures \citep[KELT-9b;][]{gaudi17},
rocky planets in the habitable zone \citep[Kepler-186f;][]{quintana14},
a gas giant planet orbiting a brown dwarf \citep[2M1207b;][]{chauvin04}, and
an Earth-mass planet around an ultra-cool dwarf \citep[OGLE-2016-BLG-1195;][]{shvartzvald17b},  
to name a few. 
These planets have been discovered using a few different detection techniques,  
and each technique has distinct capabilities and limitations. 
By far the largest number of planets have been discovered using the transit technique, and 
in particular the yield of planets from \emph{Kepler}, the first mission to 
statistically explore the population of exoplanets over a broad region of 
parameter space, was notably high \citep{coughlin16}. 
\emph{Kepler} exoplanets are on orbits similar to 
the inner planets in the Solar System 
and in many cases are more compact than that of Mercury. 
The longest-period confirmed transiting exoplanets are: 
Kepler-1647b \citep[1108 days;][]{kostov16}, 
Kepler-167e \citep[1071 days;][]{kipping16}, and
Kepler-1654b \citep[1048 days;][]{beichman18}.
The orbital periods of these planets are shorter than the orbital periods 
of all Solar System gas and ice giants. 
The lack of a large number of the long-period planets hampers our understanding 
of the formation of planetary systems as a whole and our ability to place 
the Solar System in the context of known exoplanetary systems in particular.

The main reason for this unsatisfactory situation is that different planet 
detection techniques have different sensitivities to the wide-orbit planets.  
For giant planets, the radial velocity (RV) technique is
intrinsically limited by the length of the time-baseline of 
the RV surveys themselves \citep{kane11,sahlmann16,wittenmyer17}.  
For example, only recently did \citet{blunt19} report the detection of 
a $3M_{\rm Jup}$ minimum mass object on a $74^{+43}_{-22}~{\rm yr}$ long 
highly eccentric orbit via RVs, and in this case the detection required 
a fairly fortuitous alignment of the orbit of the planet. 
In particular, the RV data taken during periastron passage of 
the planet exhibited a signal that is highly unlikely to be produced by 
other astronomical phenomenon.
The limit set by the long-term stability of the spectrographs makes detection of 
long-period Neptune-mass planets much more difficult than long-period Jupiter-mass planets: 
the RV signals are $0.5~{\rm m\,s^{-1}}$ and $9~{\rm m\,s^{-1}}$, respectively, for 
a Neptune-mass and a Jupiter-mass planet on a $10~{\rm AU}$ 
edge-on orbit around $1~M_{\odot}$ star.  
Astrometric detection of planets on relatively wide orbits (semi-major axis up to $5\textup{--}6~{\rm AU}$) 
can be done using \emph{Gaia} data 
or by combining \emph{Gaia} and \emph{Hipparcos} data \citep{perryman14,snellen18}, but
the astrometric technique is also only sensitive down to Jupiter-mass objects for most stars.  
\emph{Gaia} mission can be extended from nominal $5~{\rm yr}$ up to $10~{\rm yr}$ and 
this will increase number of detected planets by a factor of $3\textup{--}4$ \citep{perryman14}. 
Extension of \emph{Gaia} mission improves sensitivity to wider-orbit and lower-mass planets, 
but still predicted detections are smaller than the orbit of Saturn.
Combination of RV and astrometric techniques improves constraints on  
the orbital period \citep{eisnerkulkarni02,feng19}, but fundamental limitations given earlier still apply.

There are two planet detection techniques that find wide-orbit planets: direct imaging and microlensing. 
Current direct imaging surveys \citep{baron19,nielsen19} 
can detect planets with separations from $\approx5~{\rm AU}$ to thousnads of AU 
and more massive than $\approx2$ Jupiter masses. While the separation ranges 
for direct imaging and microlensing planets overlap, there are significant 
differences.
Direct imaging discovers self-luminous planets around nearby young stars, and allows 
follow-up studies of these directly detected objects, including spectroscopy
\citep{bowler16}. 
Contarary, the microlensining planets orbit old stars and 
there is no possibility for spectroscopy of these planets. 
Comparison of statical properties of direct imaging and microlensing planets 
should give constraints on planet migration.

The microlensing technique is sensitive to planetary systems that are a few 
kpc away, and mostly probe the planetary population orbiting 
the most numerous, low-mass (and hence mostly old) stars. 
The ongoing microlensing surveys are sensitive to {planet/star} mass ratios 
smaller than $10^{-3}$ even for wide-orbit planets. In fact, 
the widest-orbit microlensing planet has mass ratio of $2.4\times10^{-4}$ 
\citep[OGLE-2008-BLG-092LAb;][]{poleski14c}.  Microlensing can probe 
Neptune-mass planets, even for planets that 
have no detectable stellar host and thus may be unbound \citep{mroz18a}.

It is important to combine the constraints of both the wide-orbit and 
the free-floating planets \citep{mroz17b} in order to fully understand the formation and evolution 
of planetary systems. The bound-planet parameters that are readily measured 
for microlensing  are the mass ratio ($q$) and 
the projected separation ($s$) in units of the Einstein ring radius ($\theta_\E$). 
The microlensing planets with the widest orbits are 
OGLE-2008-BLG-092LAb \citep[$s=5.3$;][]{poleski14c}, 
OGLE-2011-BLG-0173Lb \citep[$s=4.6$;][]{poleski18b}, and
KMT-2016-BLG-1107Lb \citep[$s=3.0$;][]{hwang19} -- see discussion in \citet{poleski18b}.  
There are only a few more planets with $s>2$.
For a typical configuration, $\theta_\E$ corresponds to around $2.5~{\rm AU}$.  Hence, 
the three widest-orbit planets are at projected separations from $7$ to $15~{\rm AU}$. 
The distribution of microlensing planets as a whole has already been studied 
statistically \citep[e.g.,][]{gould10,cassan12,suzuki16,udalski18b}, 
but the statistical properties of the wide-orbit planets  
have not yet been comprehensively analyzed, 
partly due to the small number of known such systems. 

The large sample of wide-orbit planets is important for understanding formation 
of planetary systems.  We have detailed knowledge about Uranus and Neptune, 
but explaining their formation is challenging.  \citet{pollack96} showed that 
in-situ formation in standard core-accretion model cannot reproduce properties of
Uranus and Neptune.  Three major theoretical approaches to formation 
of Solar System ice giants are: 
migration from closer orbits \citep{thommes99,tsiganis05}, 
pebble accretion \citep{lambrechts14,venturini17}, and
collisions of planetary embryos \citep{izidoro15}.  
At this point, none of these theories are favoured. 

Here we present the discovery of a wide-orbit exoplanet OGLE-2012-BLG-0838Lb. 
A short-duration anomaly is observed well before the main peak of the event and 
points to an event with $s=2.1$.  The wide-orbit planet interpretation 
is confirmed by detailed modeling.  The planetary anomaly 
was found in pure survey observations by 
the Optical Gravitational Lensing Experiment \citep[OGLE;][]{udalski15b}, 
i.e., the planet detection did not depend on targeted follow-up photometry. 
This means that the planet can be included in future statistical studies of 
the wide-orbit planets. For OGLE-2012-BLG-0838, high-resolution imaging and 
satellite imaging were collected, which helps to constrain
the planet properties. 

In the next section,
we present the data collected for OGLE-2012-BLG-0838. 
We describe the model fitting in Section~\ref{sec:model}. 
In Section~\ref{sec:prop}, we analyze current constraints on 
the physical properties of the system. We summarize the paper in 
Section~\ref{sec:sum}. 

\section{Observations} \label{sec:data}

\subsection{OGLE photometry} 

OGLE is a large scale photometric survey. It is currently in its fourth phase 
(OGLE-IV) and operates a 1.3-m telescope at Las Campanas Observatory (Chile) 
that is equipped with a 32-CCD chip camera (256M pixels in total). 
The camera field of view is $1.4~{\rm deg^2}$, and the pixel scale is 
$0\farcs26$. OGLE bulge observations are performed in the $I$ band, and 
we use only these to fit the microlensing model. When the anomaly occurred, 
the field of OGLE-2012-BLG-0838 was observed once per one or two nights 
as part of bulge survey observations aiming at finding ongoing microlensing events. 
This cadence is not enough for characterizing planetary anomalies in most cases, 
but gives targets for follow-up photometric observations. 
For OGLE-2012-BLG-0838 anomaly, there is a single epoch that deviates by more than $1~{\rm mag}$ and 
four epochs that deviate by less than $0.25~{\rm mag}$. 
There are 20 OGLE fields that are observed with higher cadence. 
For the OGLE-2012-BLG-0838 field and CCD camera chip (\#32), 
the median seeing is $1\farcs46$, which 
is slightly higher than for the same chip in higher-cadence 
bulge fields ($\approx1\farcs35$).
Additional lower cadence $V$-band data taken in survey mode 
on the target exist, but do not cover the anomaly, and 
we use them only to characterize the source star. 
Photometry of the OGLE data is performed using difference image analysis 
\citep[DIA;][]{alard00,wozniak00}. We corrected the native photometric uncertainties 
following \citet{skowron16a}. 
We use data from 2012 as well as 2011, which constrain the baseline brightness. 
For a more detailed description of the OGLE survey, 
see \citet{udalski08red} and \citet{udalski15b}.

\begin{figure}
\begin{center}
\includegraphics[width=.88\textwidth,bb=8 35 537 724]{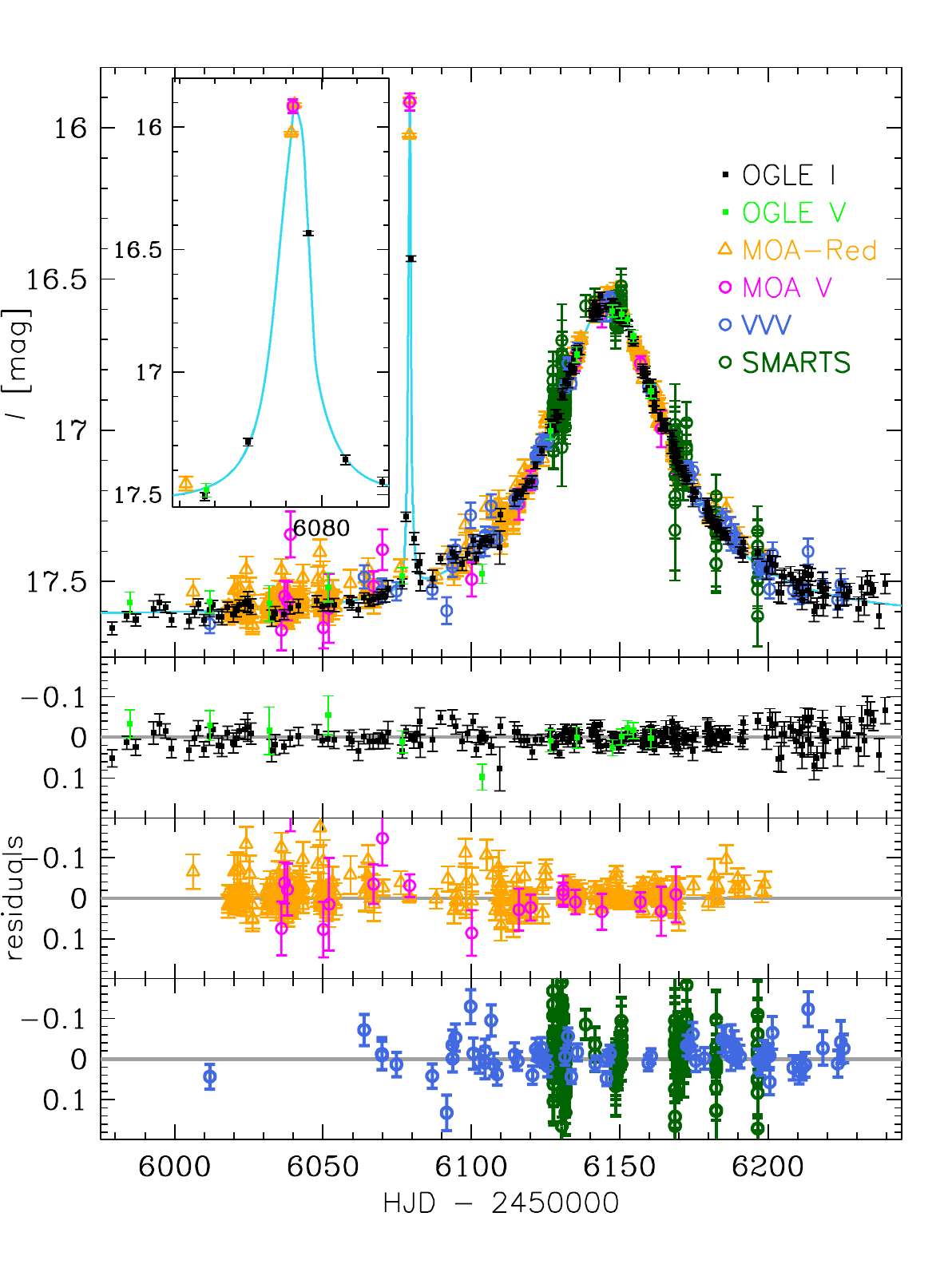}
\end{center}
\caption{Light curve of OGLE-2012-BLG-0838 and the best-fitting model top panel. 
All avaialable photometry is shown.
The inset zooms in on the anomaly.  
The lower three panels show residuals (two datasets per panel).
\label{fig:lc} 
}
\end{figure}

The search for microlensing events in the OGLE data is performed daily \citep{udalski03}.  
The event OGLE-2012-BLG-0838 was discovered on 
${\rm HJD'} \tbond {\rm HJD}-2450000 = 6082$, i.e., after the anomaly was over (see Figure~\ref{fig:lc}). 
The planetary nature of the anomaly was first suggested on 
${\rm HJD'} = 6126.403$ (by A.~U.), and subsequently the planetary models were fitted (by C.~H.).
Event coordinates are ${\rm R.A.} = 18^{\rm h}12^{\rm m}00\fs74$ and
${\rm Dec.} = -25\degr42\arcmin41\farcs8$, which translate to 
$l = 5\fdg720$ and $b = -3\fdg472$.  
The baseline brightness in the standard photometric system is:
$I = 17.610~{\rm mag}$ and $(V-I) = 1.851~{\rm mag}$ \citep{szymanski11}. 

\subsection{MOA photometry} 

The Microlensing Observations in Astrophysics \citep[MOA; ][]{bond01,sumi03b}
collaboration also conducts a microlensing survey toward the bulge by using 
the MOA-II 1.8-m telescope. The telescope is located at Mt John University Observatory 
(New Zealand). The camera used is the MOA-cam3 \citep{sako08}. It is mounted 
on the prime focus of the telescope and has field of view of $2.2~{\rm deg}^2$,
which enables high-cadence observations.  Unfortunately, the event
OGLE-2012-BLG-0838 is located in the gap between CCD chips \#3 and 
\#8 of gb18 field in the reference image. The reference images are used for the DIA pipeline 
\citep{bond01} to find and alert new microlensing events. 
Thus, the event was not discovered by the MOA collaboration and 
it was thought that MOA has no data for OGLE-2012-BLG-0838.  
The MOA data were re-checked after the initial version of this paper was submitted 
and it was revealed that the target is only $3\arcsec$ away from the edge of CCD chip \#3. 
Since the typical 
telescope pointing accuracy is larger than $3\arcsec$ and this field is observed 
in survey mode  
every 50 minutes with the custom MOA-Red filter, we were able to derive the 
MOA light curve which is dense enough to cover the anomaly. In addition to 
the MOA-Red data, occasional $V$-band data were also obtained. 
Over the anomaly, there are two MOA-Red epochs and a single $V$-band epoch and all 
these data come from a single night. 
The MOA data were reduced using the MOA's implementation of the DIA pipeline.
Also, the MOA data were corrected for the possible effects of seeing, airmass, 
and differential refraction \citep{bond17} 
by fitting to the baseline data two fifth-order polynomials of 
seeing and hour angle. The resulting correction was applied 
to all the MOA data and improved the baseline by $\Delta\chi^2=694$ for 
1344 datapoints.  The amplitude of variations are $\pm300$ 
MOA flux counts, or $\pm0.03~{\rm mag}$ at the baseline. 
Similar trends are not seen in OGLE data. 
We note that the 3 MOA data points taken during the anomaly show consistent shape of the anomaly.
In order to account for 
the underestimated uncertainties, we multiply them by $1.67$ and $1.55$ 
for MOA-Red and $V$ band, respectively.  These values were selected so that 
$\chi^2/{\rm dof} \approx 1$ for an initially fitted model.
The MOA baseline photometry shows trends on timescales on the order of one year, thus we restricted MOA data to 2012.
Similar trends were seen in previously published events and in the present case the photometry can be additionally affected by the location of the event very close to the CCD chip edge.

\subsection{\emph{EPOXI} imaging} 

Thanks to the early recognition of its anomaly, OGLE-2012-BLG-0838 was scheduled for 
observations with the \emph{EPOXI} mission, which was the repurposed \emph{Deep Impact} 
spacecraft \citep{hampton05}. There are 6516 images 
collected between ${\rm HJD'} = 6136$ and $6150$. 
The \emph{EPOXI} images are out of focus, each star produces a doughnut-shaped 
image, and the PSF changes with the color of the star.  
In the dense stellar fields of the Galactic bulge, the images of many stars 
are overlapping, which hinders photometric analysis.  
Thus the OGLE-2012-BLG-0838 \emph{EPOXI} data have not yet been reduced. 
For a reduction and analysis of the \emph{EPOXI} data for a different event, 
see \citet{muraki11}. 
Previous experience with photometric reduction of \emph{Spitzer} and \emph{K2} 
bulge images (undersampled in both cases) shows that photometric reduction of 
bulge images taken by satellite missions requires special efforts 
\citep{calchinovati15b,zhu17a,poleski19b}.

\subsection{VVV photometry} 

The Variables in the Via Lactea (VVV) survey \citep{minniti10} observed 
the Galactic bulge between 2010 and 2015 using the near-infrared 4-m VISTA telescope 
situated at the Paranal Observatory (Chile). VVV took most of its observations 
in the $K_s$ band. The event OGLE-2012-BLG-0838 is detectable in VVV data, but no 
useful data were taken during or close to the anomaly. The epoch closest to the anomaly
was secured under non-photometric conditions. Hence, the VVV $K_s$-band data 
do not usefully constrain the binary-lens microlensing model, 
and we use them only to derive the source 
properties.  Photometry was extracted using a PSF-fitting technique. 
From the VVV data we derive a baseline of $K_s = 15.190~{\rm mag}$.

\subsection{SMARTS photometry} 

Immediately following A.~U.'s planetary alert ({\rm HJD'} = 6126.403), the
Microlensing Follow Up Network ($\mu$FUN) initiated
observations using the ANDICAM dual-beam optical-IR camera
\citep{depoy03} on the
SMARTS 1.3m telescope at Cerro Tololo InterAmerican Observatory (CTIO, Chile).
The sole purpose of these observations was to characterize the source,
primarily to measure the $H$-band source flux in order to compare to
possible future high-resolution adaptive optics imaging.
During these $H$-band observations using the IR channel, the
optical channel was used to obtain $V$ and $I$ data as a backup for
the unlikely possibility of problems with the OGLE $V$-band data.
However, as anticipated, there were no such problems.  Hence, only
the $H$-band data are used in the present analysis.  Because the
observations
began before the main peak, they covered a complete range of magnifications 
of the main subevent 
from near-baseline to peak, which is the main guarantee for an accurate
measurement of the source flux.  The data were reduced using
DoPhot \citep{schechter93}. The zero-point of the photometry was
calibrated using $154$ nearby stars with VVV photometry. 
The difference between VVV photometry and SMARTS instrumental magnitudes 
shows a linear dependence on the magnitude itself and 
we take this effect into account in the zero-point calibration.  
The calibration has an uncertainty of $0.053~{\rm mag}$.  
There were a total of $205$ $H$-band observations in 10 dither or
5 dither groups
at a total of $21$ epoch, of which $150$ observations were successfully 
reduced. 
Median seeing of SMARTS data is $1\farcs2$.

\subsection{Magellan adaptive optics imaging} 

\begin{figure}
\begin{center}
\includegraphics[trim=16px 8px 16px 12px,width=0.55\textwidth]{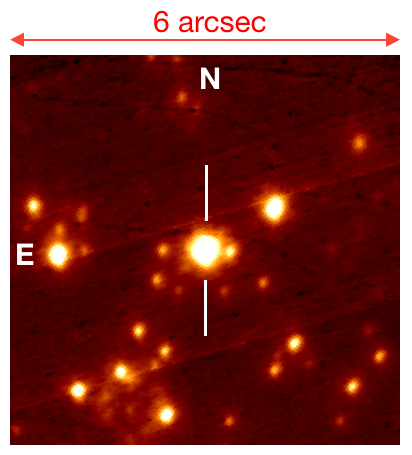}

\bigskip
\medskip
\includegraphics[trim=6px 53px 25px 14px,clip,width=0.75\textwidth]{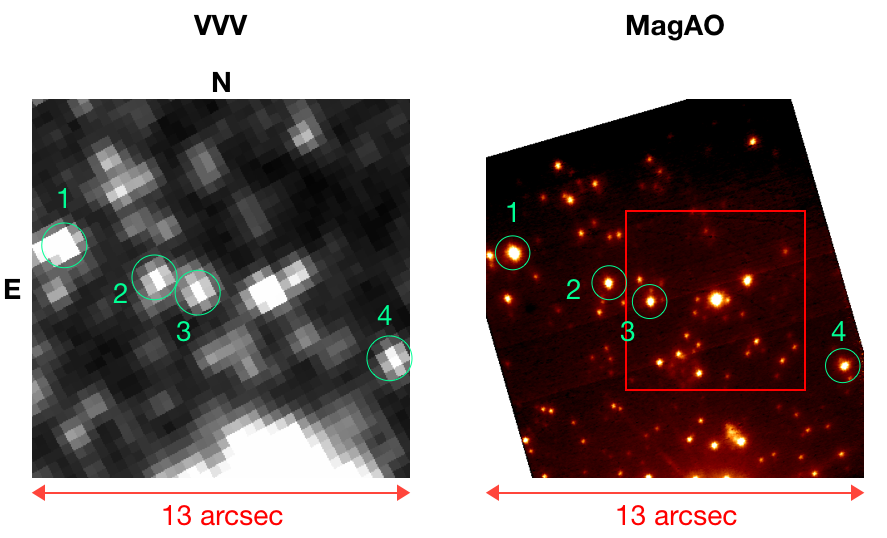}
\end{center}
\caption{MagAO image of ithe target field. 
{\it The top panel} zooms in on the target.
The flux at the location of the event is primarily due to the OGLE-2012-BLG-0838 source star.
The diagonal streaks are caused by lines of bad pixels on the individual dithered frames.
{\it The bottom panels} compare the VVV image and the MagAO image. The circles mark the four stars used for photometric calibration. The red square marks sky-area shown in top panel.
\label{fig:ao}
}
\end{figure}

The $H$-band high-resolution images of the OGLE-2012-BLG-0838 field were taken 
on ${\rm HJD'}=6766$, with Magellan Adaptive Optics system 
\citep[MagAO;][]{close12,males14,morzinski14} on the 6.5-m Clay Telescope 
at Las Camapanas Observatory (Chile). We used the Clio Wide camera, which has a plate 
scale of $27.49~{\rm mas}$ and a field of view of $14\arcsec\times28\arcsec$.
The integration time for an individual science exposure was $30~{\rm seconds}$, 
and we took ten sets of images with four dithers for each set. 
Individual dithered frames were astrometrically aligned using the positions 
of the 10 bright isolated stars, and then the aligned and 
resampled images were median-combined.
We performed the coordinate transformation from the OGLE frame 
to the MagAO frame using 
the positions of the six common isolated stars. The position of the source that 
we identify on MagAO 
image lies $(22, -14)\pm(19, 17)~{\rm mas}$ in the East and North 
relative to the transformed position of
the target centroid on the subtracted OGLE image. The closest star on the MagAO 
images is about $390~{\rm mas}$ away, so the identification of the target is 
secure (see Figure~\ref{fig:ao}).  The MagAO source is isolated with a FWHM of $160~{\rm mas}$. 
We use SExtractor \citep{bertin96} to perform aperture photometry on 
the MagAO images.  MagAO data are typically calibrated to 
the 2MASS photometric catalog \citep{skurtskie06}.  Due to the lack of overlapping 
stars between the MagAO image of OGLE-2012-BLG-0838 and 2MASS catalog, 
we used the VVV data as a bridge between 2MASS and MagAO to do 
the photometric calibration. We performed PSF photometry on the extracted 
VVV image with DoPhot \citep{schechter93} and then we used common isolated 
stars within $3\arcmin$ of the target to calibrate it to the 2MASS magnitude 
system. Only stars with $H>12.8~{\rm mag}$ are used to avoid detector 
nonlinearity for VVV.  Then we calibrated the MagAO magnitudes using four 
common isolated stars (marked in Figure~\ref{fig:ao}) between MagAO and VVV. 

\section{Microlensing models} \label{sec:model}

The light curve of OGLE-2012-BLG-0838 (Figure~\ref{fig:lc}) presents the main event, which is 
well approximated by the \citet{paczynski86} point-source point-lens model. 
The anomaly is short and high-amplitude, but its detailed shape is not well determined. 
Such events can be produced by two types of events: 
1) a binary source and a single lens \citep{gaudi98}, or
2) a single source and a binary lens.  
Furthermore, the binary-lens case presents two possibilities 
\citep[e.g.,][]{bhattacharya16,poleski18b}: 
separation can be larger or smaller than one 
(called wide and close model, respectively).  
We discuss all three possibilities below, starting from the binary lens $s>1$ 
(or wide solution), which turns out to be the correct model. 
The model fitting was performed using datasets that cover the anomaly, i.e., 
OGLE $I$-band and both MOA datasets. These data are plotted in Figure~\ref{fig:lc}.

\subsection{Wide binary-lens model} 
\label{sec:wide}

To represent a binary-lens model, we use following parameters: 
$t_0$ -- the epoch of the minimum approach,
$u_0$ -- the minimum separation (normalized to $\theta_\E$), 
$t_\E$ -- the Einstein timescale, 
$\rho$ -- the source radius (normalized to $\theta_\E$),
$\alpha$ -- the angle between the source trajectory and the lens axis,
$s$, and $q$. 
For parameter conventions we follow \citet{skowron11} and define
$t_0$ and $u_0$ relative to the primary lens. 
The first three parameters ($t_0$, $u_0$, and $t_\E$) 
are constrained by the main subevent, i.e., 
their values can be obtained by fitting a point-source point-lens model to the data with 
the short-duration anomaly epochs removed (${\rm HJD'}$ from 6077 to 6083). 
The other parameters are constrained by 
the time and length of the short-duration anomaly except $\rho$, and  
can be reasonably well estimated by visual inspection of the light curve. 
There are two additional flux parameters: the source flux and 
the blending flux. We estimate them separately for each model 
using linear regression.  The linear limb-darkening coefficients are assumed 
to be $\Gamma_I = 0.46$, $\Gamma_{MOA-R} = 0.51$, and $\Gamma_{MOA,V} = 0.66$,  
which were estimated based on a preliminary fitted model and 
the color-surface brightness relations by \citet{claret11}. 

We explored the parameter space using the Multimodal Ellipsoidal Nested Sampling
algorithm or MultiNest \citep{feroz08,feroz09}.  At each step MultiNest 
approximates the probed distribution by a union of multidimentional ellipsoids. 
MultiNest can sample the multimodal posterior and search for 
multiple separated modes, which is an important advantage.  
The search for multiple modes can be run on 
all parameters or a selected subset of parameters.  
Additionally, MultiNest properly calculates Bayesian evidence for each mode.  
In our practice, MultiNest requires more model evaluations than 
the Monte Carlo Markov Chain (MCMC) methods, but execution time is not 
a limiting factor for OGLE-2012-BLG-0838.  
MultiNest was able to model OGLE data alone, 
while MCMC methods had poor convergence.

\begin{deluxetable*}{lr}[t]
\tablecaption{Wide binary-lens model for OGLE-2012-BLG-0838 \label{tab:wide}}
\tablecolumns{2}
\tablewidth{0pt}
\tablehead{
\colhead{Parameter} & \colhead{static model} 
}
\startdata
$t_0$   & $ 6145.909 \pm 0.034 $  \\
$u_0$  & $ 0.373 \pm 0.011 $  \\
$t_{\rm E}$ (d)   & $ 40.44 \pm 0.84 $  \\
$\rho$  & $ 0.00595 \pm 0.00034 $  \\
$\alpha$ (deg)   & $ 12.66 \pm 0.11 $  \\
$s$  & $ 2.153 \pm 0.029 $  \\
$q$   & $ 0.000395 \pm 0.000033 $  \\
$F_s/F_{\rm base}$\tablenotemark{a}  & $ 0.875 \pm 0.035 $  \\
\hline 
$\chi^2/{\rm d.o.f.}$ &  $ 830.72 / 678 $ \\
\enddata
\tablenotetext{a}{$F_s$ is the source flux and $F_b$ is the blending flux. Both are for the OGLE $I$ band.}
\end{deluxetable*}

\begin{figure}
\includegraphics[width=\textwidth]{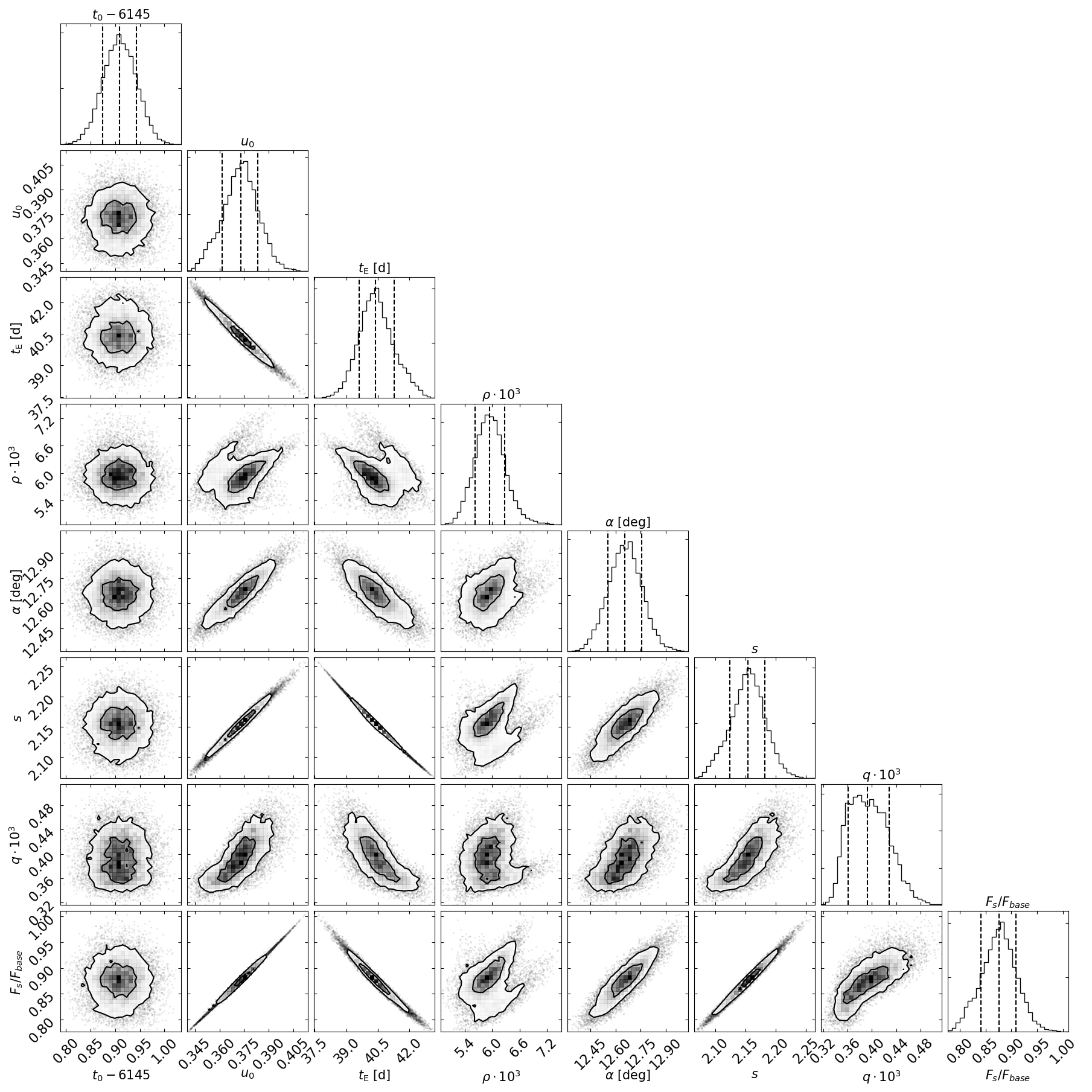}
\caption{
Marginalized posterior constraints on the microlensing parameters of the wide binary-lens model for OGLE-2012-BLG-0838. The vertical lines in 1D histograms indicate the median and $\pm1\sigma$ ranges.
\label{fig:tri}
}
\end{figure}

MultiNest found three separate modes whose main difference was 
the best-fit value of $\rho$.  In particular, MultiNest found that 
the three modes had values of
$\rho=0.0011\pm0.0007$,
$\rho=0.0037\pm0.0002$, and
$\rho=0.00595\pm0.00034$. 
The first two modes require fine tuning of 
$u_0$, $t_\E$, and $s$.  MultiNest not only searches for multiple modes 
and calculates posterior distributions of parameters, but 
also calculates the posterior probability of each mode. 
The posterior probabilities for the first and 
the second modes are smaller than the third one by a factor of $50$ and $180$, 
respectively. 
Additionally, the first mode predicts a large relative lens-source 
proper motion of $\approx16~{\rm mas\,yr^{-1}}$, which is a priori unlikely. 
Thus, the third mode is a priori preferred and we consider only 
this mode as viable in the rest of the paper 
(details of the other two models are presented in the Appendix). 
We present the results of the model fitting in
Table~\ref{tab:wide} and 
Figure~\ref{fig:tri}. The best model is plotted in 
Figure~\ref{fig:lc}.  

\begin{figure}
\centering
\includegraphics[bb=32 28 540 717,angle=270,width=.6\textwidth]{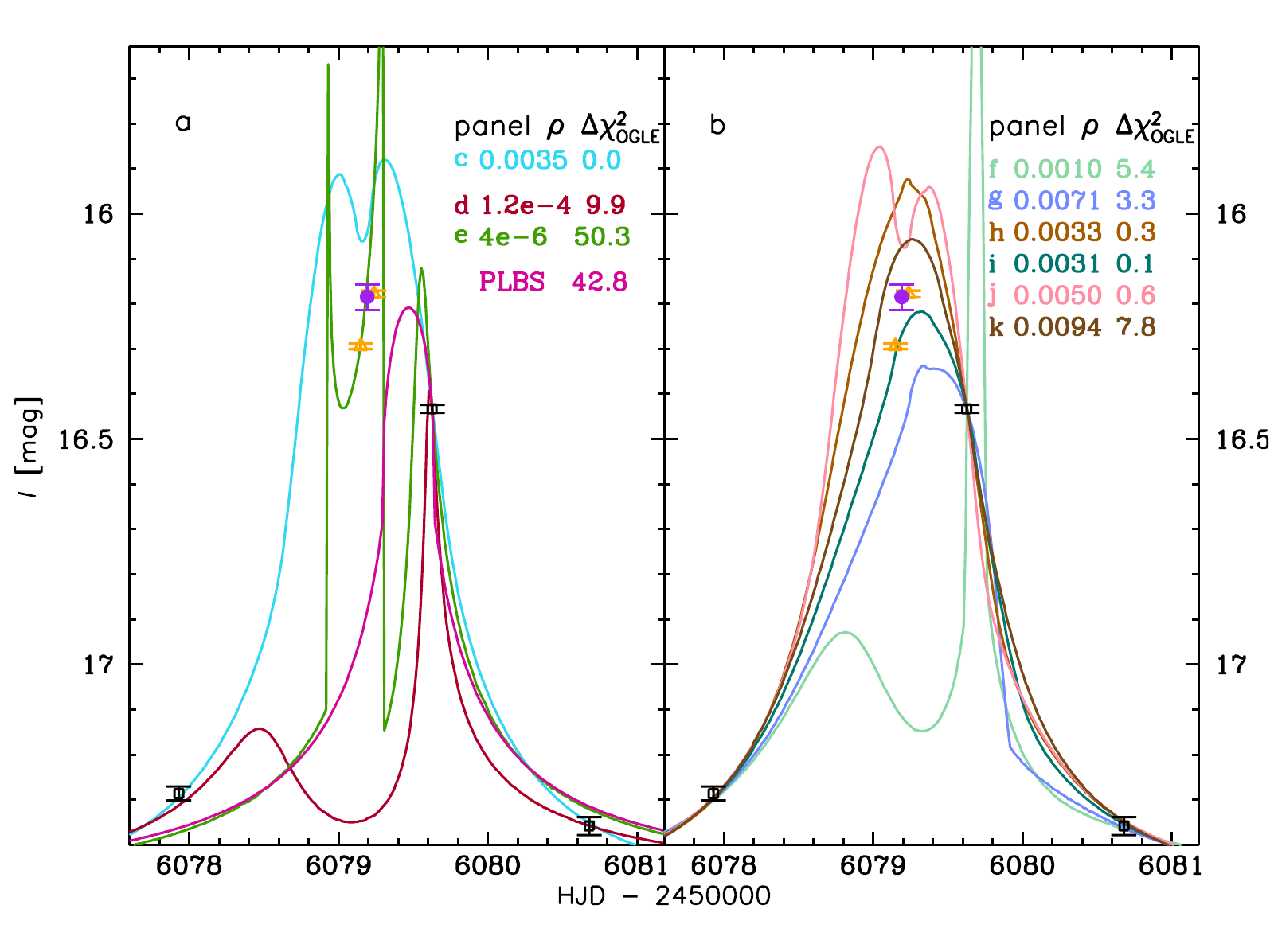}

\bigskip
\includegraphics[bb=2 0 544 545,clip,width=.465\textwidth]{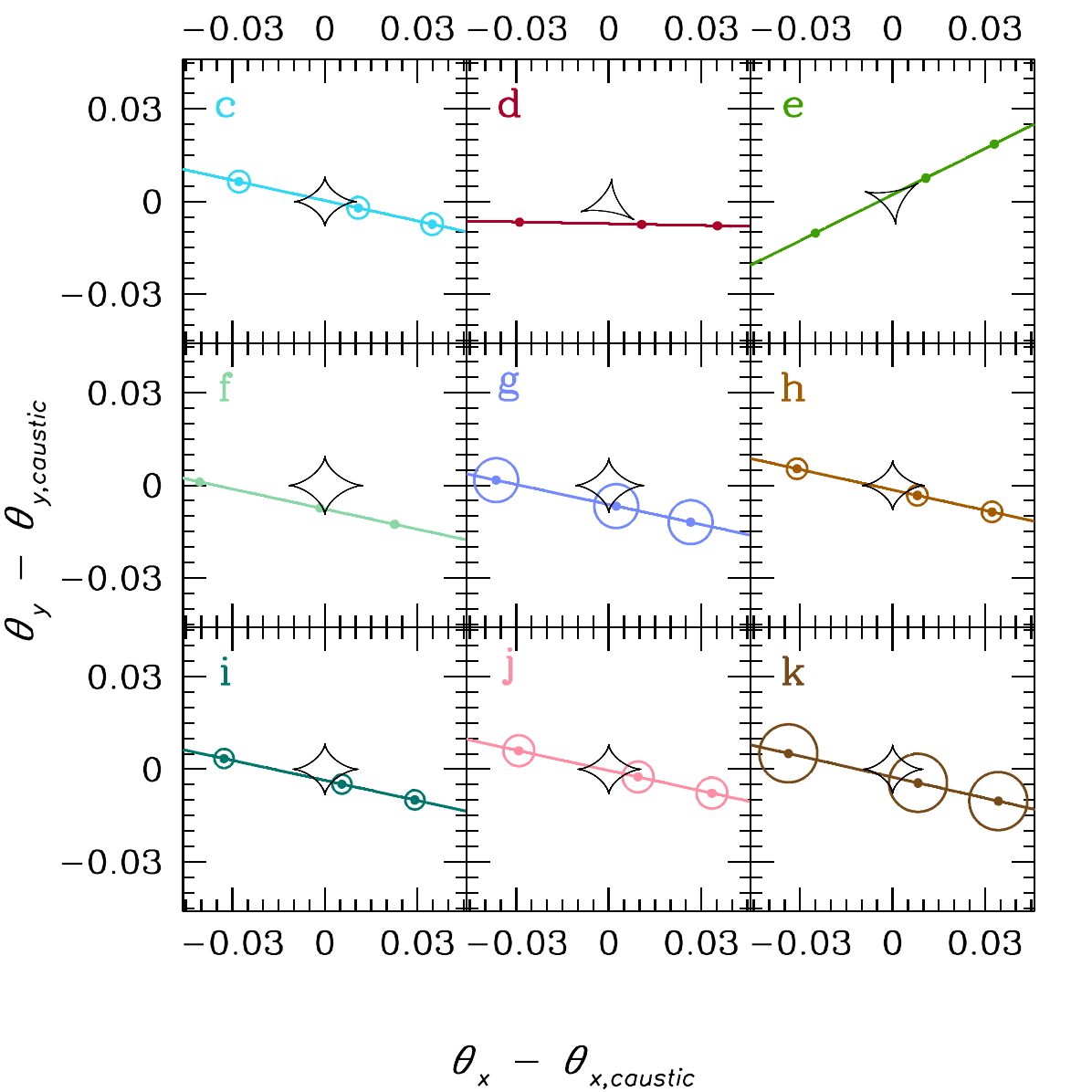}
\caption{Degenerate microlensing models for OGLE-2012-BLG-0838 fitted to OGLE data only. 
{\it The upper panels (a and b)} present model light curves, three OGLE epochs 
that constrain the anomalous part of the model 
(black; at $6077.9$, $6079.6$, and $6080.7$), 
and three MOA epochs (orange and purple; between $6079.15$ and $6079.24$).
The lines shown in panel {\it a} represent best models for each of four topologies
(wide: {\it c} and {\it f-k}; two close: {\it d} and {\it e}; and binary-source: pink line in panel {\it a}). 
The lines shown in panel {\it b} are all wide binary-lens models 
and were selected from the search for multiple modes run only on $\rho$.  
The legend gives $\rho$ and $\Delta\chi^2_{\rm OGLE}$ values for each model. 
Two models peak beyond the plot at $(6079.3,\,13.55)$ and $(6079.7,\,15.20)$.
{\it The lower panels (c-k)} show the corresponding trajectories and 
planetary caustics (black) for the binary-lens models.
The colored circles represent the size of the source as well as its position 
at the times when the three OGLE epochs were taken. 
The source is moving from left to right. 
The coordinate system is centered on a planetary caustic. 
In {\it panels d and e}, the two triangular caustics correspond to close models. 
In this coordinate systems, the central caustics are at $(1.56,\,-0.38)$ and $(1.57,\,0.38)$, respectively. 
For the other models (i.e., wide), the primary is at $(\approx 1.6,\,0)$. 
For the three models with $\rho<0.002$ {\it (panels d, e, and f)}, 
the actual source size is smaller than the points shown.
\label{fig:degen}
}
\end{figure}

It turns out that the combination of the MOA and OGLE data restricts 
the number of separate modes significantly better than the OGLE data alone. 
When we fitted only the OGLE data and searched for multiple modes on all parameters, 
then MultiNest reported only a single mode. When the search for multiple modes 
was run only on $\rho$, then 10-30 modes were found, depending on the exact settings. 
The $1\sigma$ ranges of the posterior parameters of these modes were overlapping, 
which showed that the OGLE data alone do not allow a unique identification of 
multiple modes.
We inspected many modes and in Figure~\ref{fig:degen} present 
a few modes, which were selected to show the whole range of the diversity of the light curves. 
The problems we faced with fitting the OGLE data alone 
are a less-severe case of the discrete and continuous degeneracies seen in the case of 
OGLE-2002-BLG-055 \citep{gaudi04}, which also had only a single epoch that is 
much brighter than predicted by the point-source point-lens model.

We derive the source brightness using posterior distributions. 
We obtain 
$V_s = 19.596 \pm 0.044~{\rm mag}$,
$I_s = 17.754 \pm 0.043~{\rm mag}$, 
$H_s = 15.484 \pm 0.043~{\rm mag}$, and 
$K_{s,s} = 15.326 \pm 0.044~{\rm mag}$. 
We also use calibrations of the MOA photometry to the OGLE-III magnitude system 
\citep{szymanski11} to derive the source brightness from the MOA data. The 
transformations have a scatter of $0.048~{\rm mag}$ in the $V$ band and 
$0.045~{\rm mag}$ in the $I$ band. We obtain $V^{\rm MOA}_s = 19.587\pm0.043~{\rm mag}$ 
and $I^{\rm MOA}_s = 17.662\pm0.043~{\rm mag}$, which are consistent with $V_s$ and $I_s$
derived from the OGLE data within uncertainties.

After considering the static binary-lens model, we attempted to include 
the microlensing parallax effect.  Microlensing parallax is described by a 2D vector 
$\pmb{\pi_\E}$, whose amplitude is equal to the relative lens-source parallax 
divided by $\theta_\E$. If both $\theta_\E$ and $\pmb{\pi_\E}$ are measured, 
then both the lens mass ($M$) and distance ($D_l$) 
are measured directly \citep{gould00b}:
\begin{equation}
M=\frac{\theta_\E}{\kappa\pi_\E},~~~~~~
\frac{1}{D_l} = \frac{\pi_\E\theta_\E}{\rm AU} + \frac{1}{D_s},
\end{equation}
where $\kappa = 4{\rm G}/({\rm AU}\,{\rm c}^2) = 8.14~{\rm mas~M_{\odot}^{-1}}$ is a constant, 
and $D_s$ is the source distance. 
The annual microlensing parallax breaks the assumption 
that the apparent lens-source relative motion is rectilinear. 
The effect is undetectable for most events, because during their (typically short) duration,
Earth's motion around the Sun can be well approximated by a straight line. 
OGLE-2012-BLG-0838 has relatively long $t_\E$ of $40$ days. 
The anomaly additionally increases the chances of measuring $\pi_\E$, because 
it provides a well-timed event \citep{an01}.  

We consider two degenerate scenarios: $u_0<0$ and $u_0>0$. 
The best-fitting parallax model improves $\chi^2$ by $23.4$ and the uncertainty of
$\pi_{{\rm E}, N}$ is large ($\sigma_{\pi_{{\rm E}, N}} = 0.33$). 
We checked a plot of $\chi^2$ difference between the best parallax model 
and the best static model. It revealed that there may be a problem with the quality 
of the MOA data on a timescale of dozens of days. It is known that low-level systematics
may mimic the microlensing parallax signal. Some trends in residuals of static
binary-lens model can be seen in Figure~\ref{fig:lc}, e.g., around ${\rm HJD'} = 6100$. 
Thus, we decided to report only 
well-established static model and do not present potentially spurious parallax models.

\subsection{Close binary-lens model} 

\begin{figure}
\includegraphics[width=\textwidth]{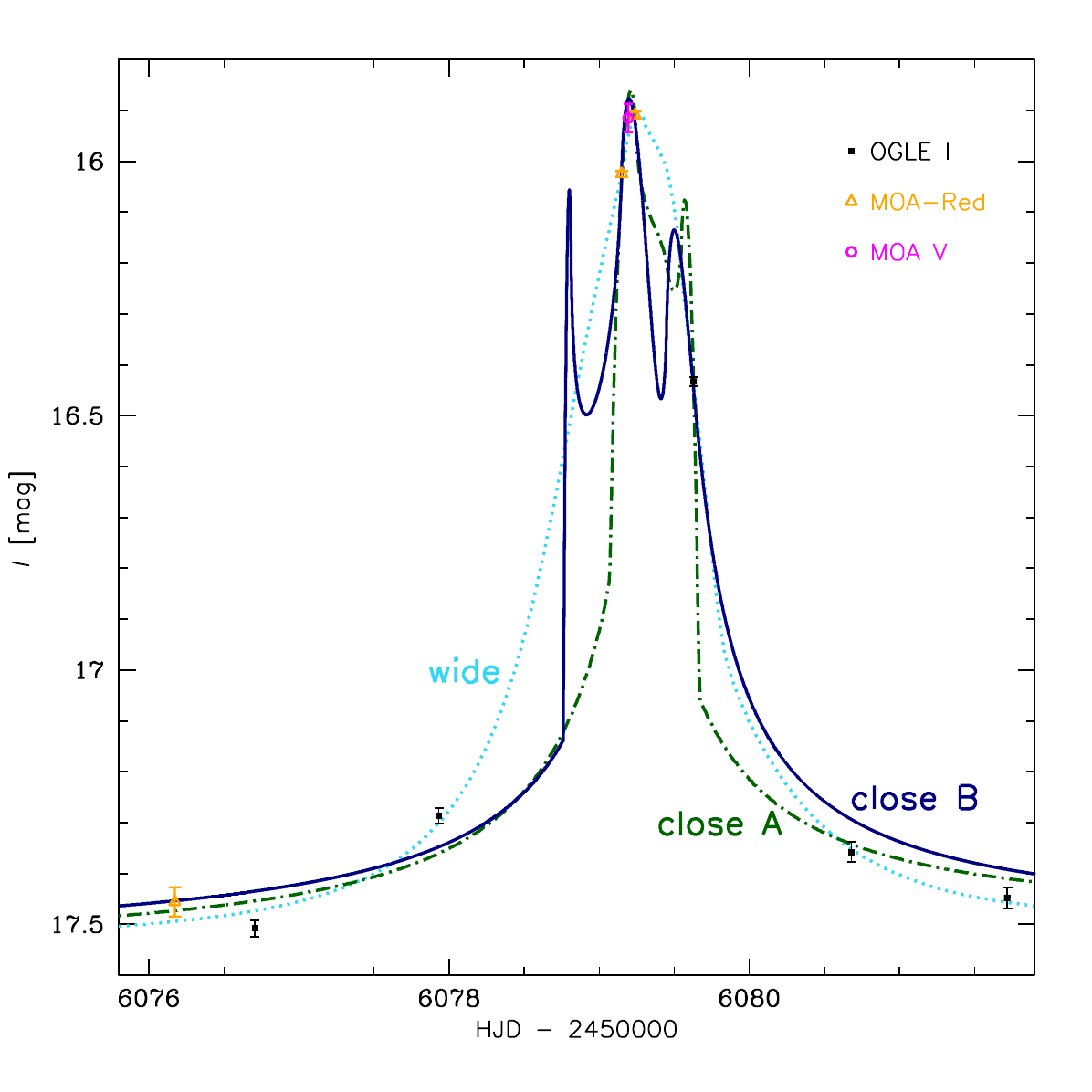}
\caption{Three binary-lens models for the anomalous part of the light curve fitted to OGLE and MOA data. 
There is a single wide model (turquoise dotted line; same as in Figure~\ref{fig:lc}) and two close models:
with $\alpha = 177~{\rm deg}$ (marked A, dark green dotted-dashed line) and 
with $\alpha = 207~{\rm deg}$ (marked B, dark blue solid line).
\label{fig:lcA}
}
\end{figure}

We additionally searched for close (i.e., $s<1$) binary-lens models.
and found two such solutions.  The parameters of these models are presented in the Appendix. 
The first model (marked ``close A'' in Figure~\ref{fig:lcA}) has
$\chi^2 = 857.84$, i.e., it is worse fit to the data by $\Delta\chi^2=27.12$. 
The wide model is favored over the close model a priori.
First, the wide model predicts the relative lens-source proper motion of 
$\approx3~{\rm mas\,yr^{-1}}$ (see below), which is the typical value, 
while the close model predicts much less likely $\approx15~{\rm mas\,yr^{-1}}$. 
Second, a recent statistical analysis of microlensing events \citep{suzuki16} 
shows that the microlensing planet occurrence rate is increasing with increasing $s$ 
and decreasing $q$, and this result is confirmed by a joint analysis of microlensing, 
radial velocity, and direct imaging results \citep{clanton16}.  
We reject the close model based on $\Delta\chi^2$, 
as well as the two arguments given above.

The close model with $\alpha = 177~{\rm deg}$ has a source trajectory that 
crosses the binary axis outside the caustics 
(in other words, the source passes all caustics on the same side). 
There is a second model in which the source trajectory crosses 
the binary axis between the planetary and central caustics 
\citep[see the lower part of Figure~3 in][]{poleski18b}. 
For OGLE-2012-BLG-0838, the latter model has $\alpha=207.20\pm0.98~{\rm deg}$ 
(marked ``close B'' in Figure~\ref{fig:lcA})
and a corresponding $\chi^2$ is $876.34$, i.e., which is sufficiently larger 
($\Delta\chi^2 \simeq 45.62$) than the best fit (wide model) $\chi^2$ to be rejected. 
We compare all three binary-lens models for 
the anomaly part of the light curve in Figure~\ref{fig:lcA}.

\subsection{Binary-source model} 

The binary-source model introduces three additional parameters as compared to 
the point-source point-lens model ($t_{0,2}$, $u_{0,2}$ and flux ratio of two sources).  
The best binary-source model has $\chi^2$ of $988.98$, i.e., worse by $\Delta\chi^2 \simeq 158$ than 
the wide binary-lens model (see the Appendix). 
Clearly, the wide binary-lens model fits 
the data better, and we reject the binary-source model.

\section{System properties} \label{sec:prop}

Here we discuss a few different pieces of information about the lens and source. 
We are not able to directly measure the lens mass and distance, but we discuss the 
prospects for doing so. Thus, we derive 
the lens properties using Bayesian priors derived using a Galactic simulation. 
We also discuss the origin of the excess flux.

In Figure~\ref{fig:ao}, we show the MagAO image of OGLE-2016-BLG-0838. 
The final calibrated $H$-band brightness of the target is 
${H_{\rm target}}=15.29\pm0.05~{\rm mag}$,  
where the uncertainty estimate combines the statistical and systematic components. 
$H_{\rm target}$ is brighter than the $H$-band source flux measured using SMARTS photometry and 
the difference corresponds to $H_{\rm excess} = 17.26^{+0.46}_{-0.33}~{\rm mag}$. 
Later in this section we discuss where this excess flux comes from.

The relative lens-source proper motion is: 
$\mu_{\rm rel} = \theta_\E/t_\E = 3~{\rm mas\,yr^{-1}}$ (see below). 
We may expect that the lens and source could be resolved in about ten years from now
allowing the lens flux to be measured and leading to an estimate of 
the lens mass and distance, when combined with the stellar isochrones 
\citep{yee15c} and the constraint on the angular Einstein ring radius 
from the detection of finite source effects.
In some cases, an identification of the lens in the follow-up high-resolution imaging 
is problematic \citep{bhattacharya17}.  
The future lens flux measurement can definitely 
use the MagAO image presented here for calibration.  
We also list nearby stars in Table~\ref{tab:close}.  
As one can see in Figure~\ref{fig:ao}, the event is by far the brightest object 
within the ground-based seeing limit.  

\begin{deluxetable*}{rrrrr}[t]
\tablecaption{Stars detected close to the target on MagAO image\label{tab:close}}
\tablecolumns{5}
\tablewidth{0pt}
\tablehead{
\colhead{No.} & \colhead{distance} & \colhead{$\Delta \alpha\cos\delta$} & \colhead{$\Delta\delta$} & \colhead{$H~{\rm [mag]}$} \\
\colhead{} & \colhead{[arcsec]} & \colhead{[arcsec]} & \colhead{[arcsec]} & \colhead{} 
}
\startdata
1 & $0.39$ & $-0.379$ & $-0.026$ & $17.474\pm0.060$ \\
2 & $0.64$ &  $0.644$ &  $0.003$ & $18.073\pm0.084$ \\
3 & $0.87$ &  $0.742$ & $-0.449$ & $18.174\pm0.070$ \\
4 & $1.02$ & $-0.884$ & $-0.501$ & $18.469\pm0.073$ \\
5 & $1.24$ & $-1.056$ &  $0.652$ & $16.307\pm0.053$ \\
\enddata
\tablecomments{$\Delta \alpha\cos\delta$ and $\Delta\delta$ indicate the displacement from 
the target along R.A. and Dec. directions (i.e., E and N have positive values), respectively.}
\end{deluxetable*}

\begin{figure}
\plotone{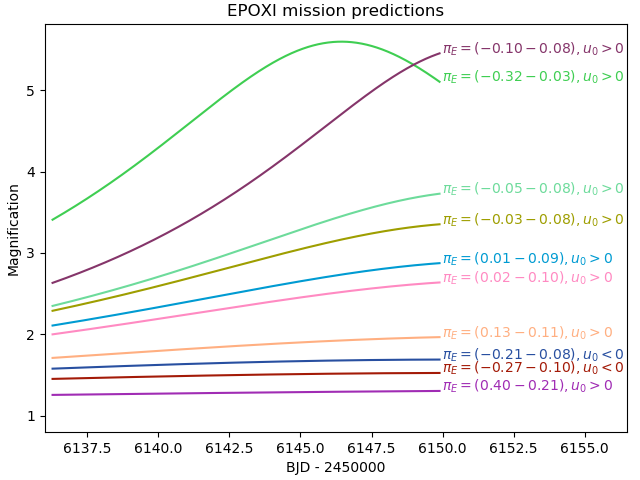}
\caption{Representative magnification curves predicted for \emph{EPOXI}. The models are consistent with the OGLE data and show a range of possible magnification curves. 
\label{fig:epoxi}
}
\end{figure}

\begin{figure}
\plotone{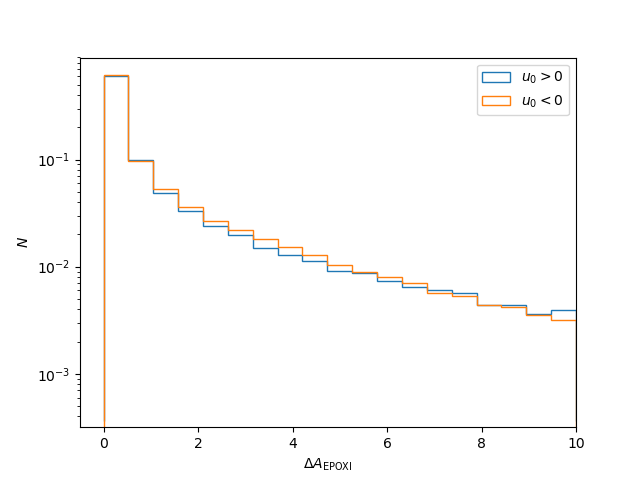}
\caption{Histogram of the predicted magnification amplitude for \emph{EPOXI}. 
The $\Delta A_{\rm EPOXI}$ values larger than $10$ are not excluded but 
have very low probability. 
\label{fig:epoxiH}
}
\end{figure}

The existing \emph{EPOXI} data have not yet been reduced. 
We use representative parallax wide binary-lens models (all are witihin $2\sigma$) fitted to the OGLE data to predict  
the magnification as seen by \emph{EPOXI} --- see Figure~\ref{fig:epoxi}. 
We also show a histogram of the amplitude (i.e., difference between maximum and minimum) 
of magnification predicted for \emph{EPOXI} in Figure~\ref{fig:epoxiH}.  
The lens mass and distance can be measured directly if 
the microlensing parallax is measured.  Some of the magnification 
curves are almost flat.  If the true magnification curve is almost flat, 
then the parallax measurement is unlikely.  If the highest magnification 
is $\lesssim4$, then the magnification curve can be reasonably well approximated 
as a linear function of time.  In this case, 
it will be necessary to remove potential 
systematic linear trends in the \emph{EPOXI} photometry in a manner that is independent of the
photometry of the source in the \emph{EPOXI} data
in order to measure $\pmb{\pi_\E}$. 

\begin{figure}
\includegraphics[width=\textwidth]{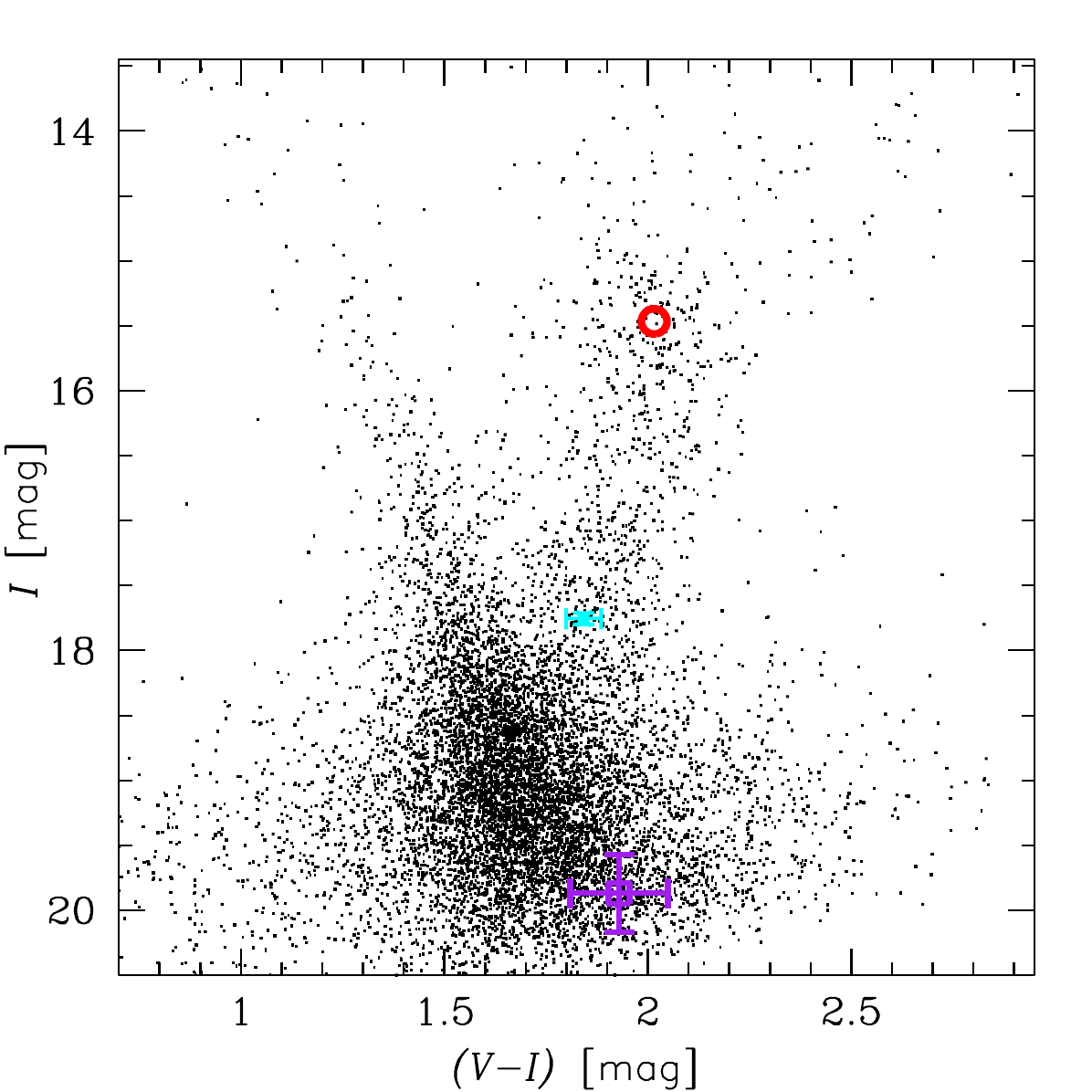}
\caption{Color--magnitude diagram for stars within $2\arcmin$ around the target. 
The cross marks the source position as derived from the posterior distribution. 
The red circle indicates the red clump (uncertainty not shown because it is smaller than the point size) and the purple square marks 
the blending light ($I_b=19.87\pm0.30~{\rm mag}$, ${(V-I)}_b=1.93\pm0.12~{\rm mag}$).  
The plotted OGLE data were presented and calibrated to standard photometric system by \citep{szymanski11}.
\label{fig:cmd}
}
\end{figure}

The event parameters could be better constrained if the proper motion of the source was known. 
The baseline object is included in \emph{Gaia} DR2 \citep{gaia18a}.  
The astrometric $\chi^2/{\rm dof}$ (keyword {\tt astrometric\_gof\_al}) is $3.9$ 
while values $>3$ indicate bad fit to the data.  The interpretation of 
the \emph{Gaia} proper-motion measurement is additionally hindered by 
the fact that the baseline object is a blend of sources with detectable contribution 
from other stars (see Table~\ref{tab:wide}).  It is unlikely that \emph{Gaia}  DR2 
proper motion can put useful constraints on system properties.

To measure $\theta_\E$, we use the method developed by \citet{yoo04b}. 
We present the color-magnitude diagram of stars lying within $2\arcmin$ from the target 
in Figure~\ref{fig:cmd}. The red clump has an observed color of $(V-I)_{\rm RC} = 2.003\pm0.008~{\rm mag}$ 
and a brightness of $I_{\rm RC} = 15.489\pm0.030~{\rm mag}$. We compare these values with 
the extinction-corrected values from \citet{bensby11} and \citet{nataf13b}, 
respectively, to obtain $E(V-I) = 0.943~{\rm mag}$ and $A_I = 1.205~{\rm mag}$. 
The extinction-corrected source properties are 
$I_{s,0} = 16.550\pm0.043~{\rm mag}$ and $V_{s,0} = 17.450\pm0.066~{\rm mag}$ 
and using the \citet{bessell88} color-color relations we obtain 
$(V-K)_{s,0}=2.022\pm0.147~{\rm mag}$. 
The estimated $(V-K)_{s,0}$ and $V_{s,0}$ correspond to 
$\theta_{\star} = 1.93\pm0.14~{\rm \mu as}$ 
\citep{kervella04b}. When combined with $\rho$ for the wide model 
we obtain $\theta_\E = 0.325\pm0.029~{\rm mas}$. 

\begin{figure}
\begin{center}
\includegraphics[width=0.8\textwidth]{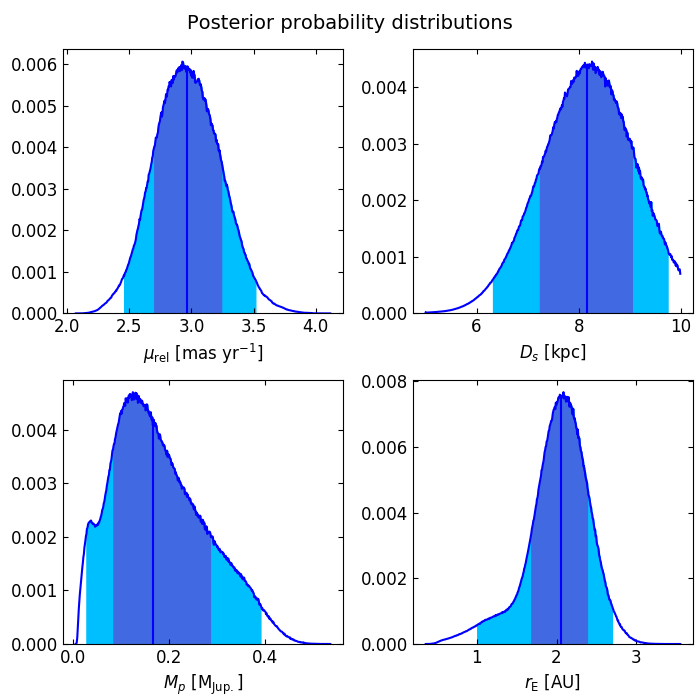}
\end{center}
\caption{
Posterior probability distributions from the Bayesian simulations of the Galaxy. 
Vertical lines indicate median values. Shade regions mark $\pm1\sigma$ and 
$\pm2\sigma$ ranges. 
Posteriors for lens mass and distance are plotted in Figure~\ref{fig:naoki1}.
\label{fig:galpost}
}
\end{figure}

We do not have an interesting constraint on the microlensing parallax.  
Hence, we must use the Bayesian simulations of the Galaxy to derive 
the lens mass and distance. For this purpose, 
we use an approach similar to that presented by \citet{clanton14a}. 
The lenses are drawn from the density profiles of 
a double-exponential disk \citep{zheng01} and boxy Gaussian bulge \citep[model G2 by][]{dwek95}, 
which are normalized according to \citet{gould96c} and \citet{hangould03}, respectively.
The lens mass function is taken from the model 1 in \citet{sumi11}. 
For the source distance we use the boxy Gaussian bulge distribution, i.e., 
model G2 by \citet{dwek95} and weight it by $D_s^2$. 
The kinematics of the disk and bulge follow \citet{clanton14a}.
The event rate $\Gamma$ is weighted according to \citep{clanton14a}: 
\begin{equation}
\frac{d^4\Gamma}{dD_l dM_l d^2\pmb{\mu_{\rm rel}}} = 2r_\E \mu_{\rm rel}D_l\nu \frac{d^2\Gamma}{d^2\pmb{\mu_{\rm rel}}}\frac{d\Gamma}{dM_l},
\end{equation}
Where $r_\E=D_l\theta_\E$ is the Einstein ring radius projected on the lens plane and
$\nu$ is the position-dependent density of lenses.
A total of $1.5\cdot10^9$ events were drawn.
The results of the simulations are presented in Table~\ref{tab:gal} 
and in Figure~\ref{fig:galpost}. 

\begin{deluxetable*}{llr}[t]
\tablecaption{Posterior Physical Parameters Statistics \label{tab:gal}}
\tablecolumns{3}
\tablewidth{0pt}
\tablehead{
\colhead{Parameter} & \colhead{unit} & \colhead{value}  
}
\startdata
$\theta_E$ &  mas &  $0.325\pm0.029$ \\
$\mu_{\rm rel}$ & ${\rm mas\,yr^{-1}}$ &
                            $2.96\pm0.27$ \\
$D_s$ & kpc &               $8.17\pm0.91$ \\
$D_l$ & kpc &               $6.32_{-1.04}^{+0.83}$ \\
$M_l$ & ${\rm M_{\odot}}$ & $0.40_{-0.20}^{+0.29}$ \\
$M_p$ & ${\rm M_{Jup.}}$ &  $0.167_{-0.83}^{+0.121}$ \\
$r_\E$ & AU &               $2.06_{-0.38}^{+0.33}$ \\
$r_{\perp}$\tablenotemark{a} & AU & $4.43_{-0.81}^{+0.71}$
\enddata
\tablenotetext{a}{Instantaneous projected star-planet separation: $r_{\perp}=s D_l \theta_{\rm E}$.}
\end{deluxetable*}

\begin{figure}
\begin{center}
\includegraphics[bb=40 39 570 743,angle=270,width=.685\textwidth]{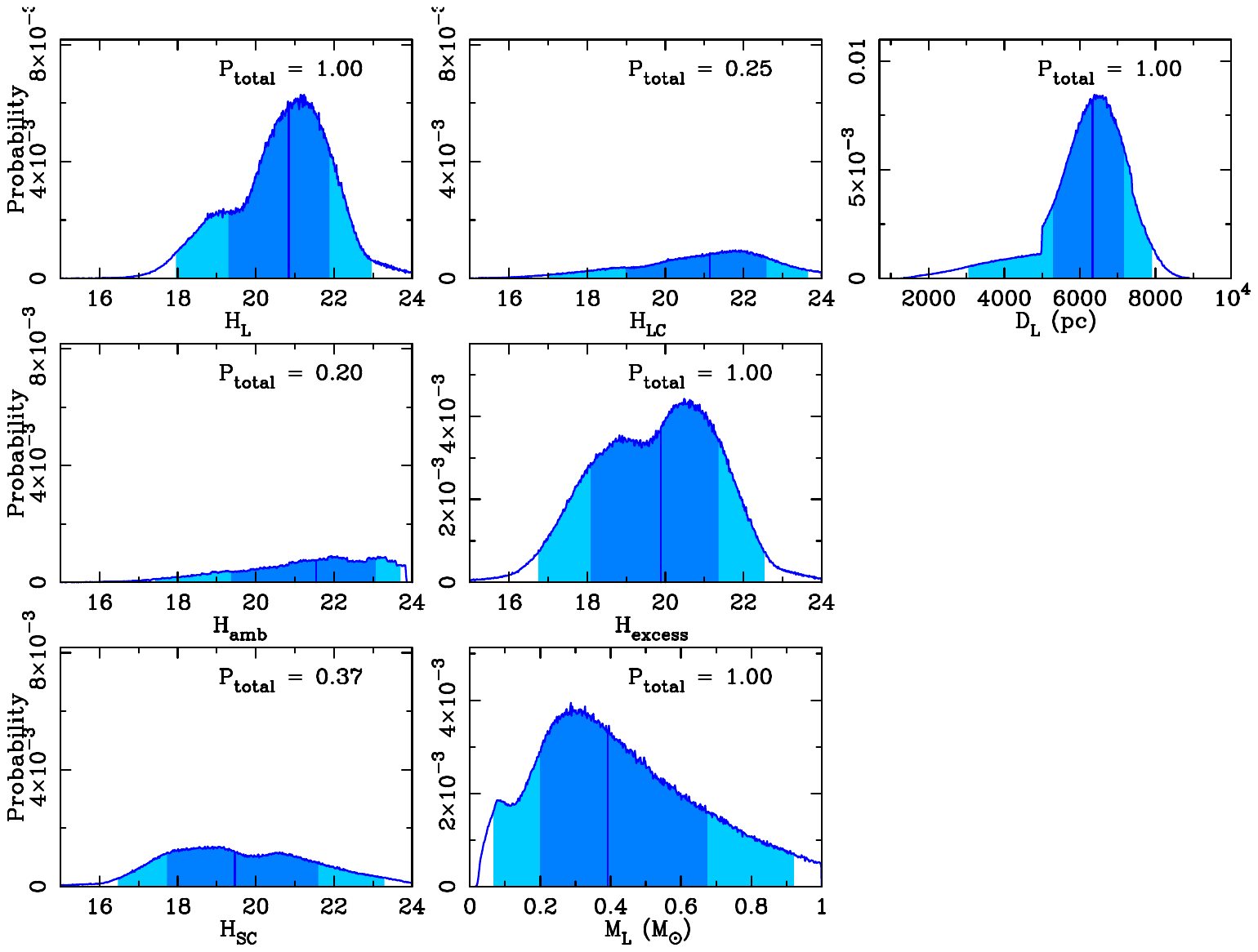}
\end{center}
\caption{
Prior probability distributions (i.e., before applying excess flux constraint). 
The dark vertical lines indicate median values and shaded regions mark
$\pm1\sigma$ and $\pm2\sigma$ ranges. Each panel gives the probability that 
given object exists. Subscripts SC, amb, and LC stand for 
source companion, ambient star, and lens companion, respectively. 
\label{fig:naoki1}
}
\begin{center}
\includegraphics[bb=40 39 570 743,angle=270,width=.685\textwidth]{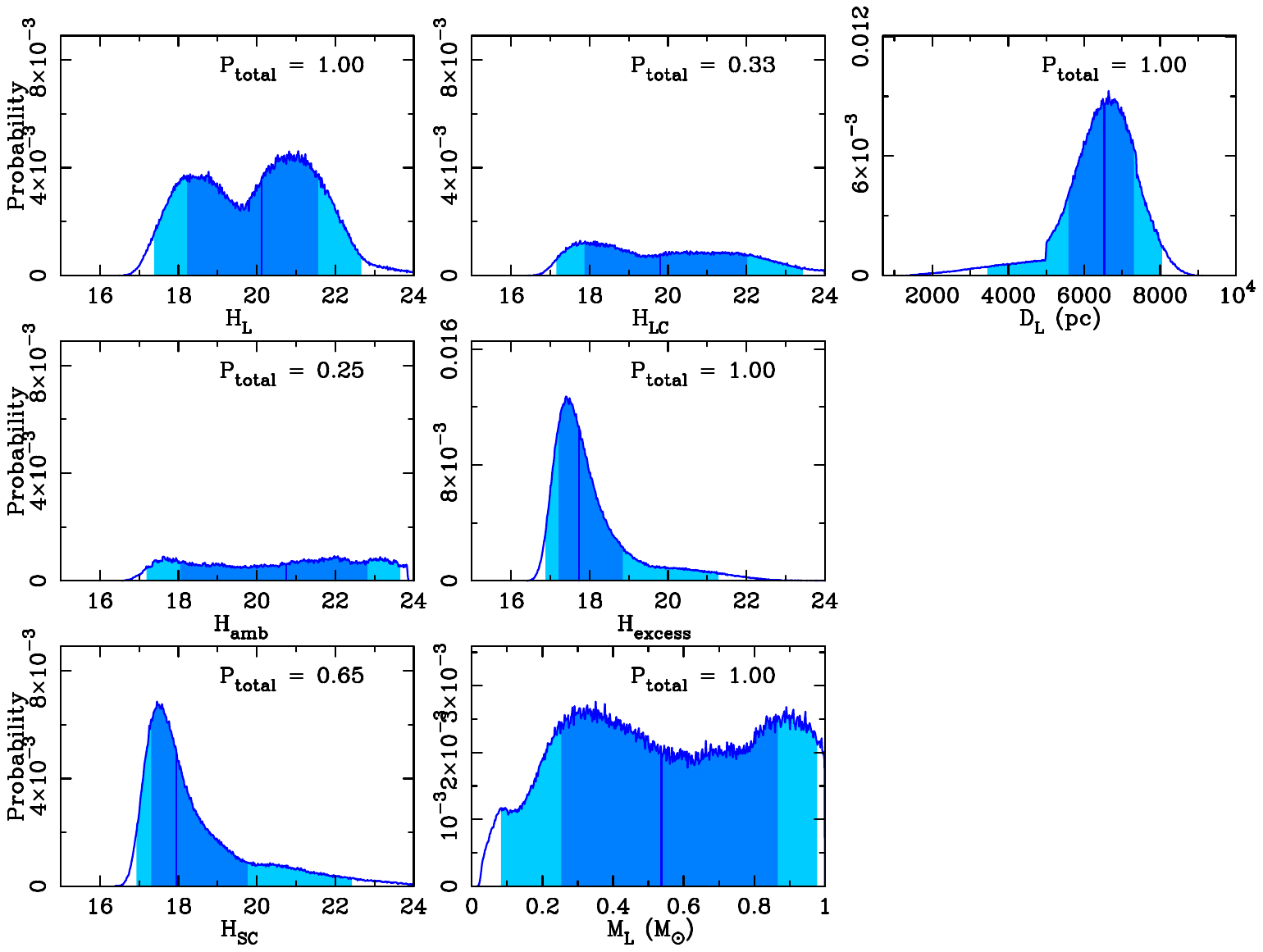}
\end{center}
\caption{
Posterior probability distributions after the $H$-band excess flux constraint is
applied to prior distributions presented in Figure~\ref{fig:naoki1}.
\label{fig:naoki2}
}
\end{figure}

There are four possible sources contributing to the measured excess flux: 
the lens, unrelated ambient star(s), a companion to the source star, 
and a companion to the lens star. 
Following the method developed by \citet{koshimoto17} and fully described by \citet{koshimoto19tmp}, we calculate 
the probabilities of all possible combinations of the four sources 
that explain the observed excess flux, assuming that none of them is a stellar remnant. 
We use posterior distributions of $D_s$, $D_l$, 
and $M_l$ from the above Bayesian simulations as a prior to analyze the origin of the excess flux.
We use the luminosity function from \citet{zoccali03} which is normalized 
to the stellar density in the OGLE-2012-BLG-0838 field to calculate 
the ambient stars flux distribution, where the normalization is done by 
comparing the number of the red clump stars in this field to that in 
the \citet{zoccali03} field using the OGLE-III catalog \citep{szymanski11}.
For the flux distribution of a companion to the source or the lens star, 
we use the binary distribution, which is based on \citet{wardduong15} 
and on the summary in 
a review paper of stellar multiplicity by \citet{duchene13}.
We consider only companions to the source or lens whose mass ratio 
and semimajor axis are consistent with both the light curve and 
the MagAO image where no signal of a detectable stellar companion is shown. 
The prior distributions of parameters ($M_l$, $D_l$, and $H$-band magnitudes) 
are shown in Figure~\ref{fig:naoki1}.

After deriving the prior distributions, we apply an $H_{\rm excess}$ 
constrained as detailed by \citet{koshimoto19tmp}.
We present the resulting posterior distributions in Figure~\ref{fig:naoki2}. 
The measured $H_{\rm excess}$ is brighter than in prior 
(see center panel in Figure~\ref{fig:naoki1}), hence, adding an $H_{\rm excess}$ constraint 
increases the estimate of the lens mass 
(from $0.40_{-0.20}^{+0.29}~{\rm M_{\odot}}$ to $0.54_{-0.29}^{+0.33}~{\rm M_{\odot}}$).  
Also the posterior probabilities that the lens companion, the source companion, 
or the ambient star contribute to $H_{\rm excess}$ are higher than 
the corresponding prior probabilities. In particular, 
the probability that the source companion contributes to $H_{\rm excess}$ increased from $0.37$ to $0.65$.  
The main origin of $H_{\rm excess}$ is therefore more likely to be the lens or the source companion. 

We can estimate the expected RV signal from OGLE-2012-BLG-0838Lb.  
We assume the median values for stellar host scenario from Table~\ref{tab:gal}, 
i.e., $M_l=0.40~{\rm M_{\odot}}$ and $r_{\perp} = 4.43~{\rm AU}$.
We estimate the semi-major axis 
assuming a random position of the planet and a circular orbit, 
$a = (3/2)^{1/2}r_{\perp} = 5.4~{\rm AU}$.  
The orbital period is 
$P= (a^3/M)^{1/2} = 19.9~{\rm yr}$.  For the edge-on configuration,
the RV signal would be $K=3.2~{\rm m\,s^{-1}}$.  
Detecting planets with similar properties around nearby stars would be 
challenging for the RV surveys.  The longest-period RV planets with 
well-measured RV curves are 
HR~5183b \citep[$P\approx74~{\rm yr}$ and $K=38.3~{\rm m\,s^{-1}}$;][]{blunt19},
HD~30177c 
\citep[$P=20.8~{\rm yr}$ and $K=35.8~{\rm m\,s^{-1}}$ or $P=31.8~{\rm yr}$ and $K=59.4~{\rm m\,s^{-1}}$;][]{wittenmyer17}
and GJ~676Ac \citep[$P=20.4~{\rm yr}$, $K=90.0~{\rm m\,s^{-1}}$;][]{sahlmann16},  
though for neither of them the RV data cover the full orbital period. 
The amplitudes of the RV signals for these three planets are more than 
an order of magnitude larger than predicted for OGLE-2012-BLG-0838Lb.

\section{Summary} \label{sec:sum}

We have presented the microlensing discovery of a wide-orbit planet OGLE-2012-BLG-0838Lb. 
Alternative models of observed light curve were considered and found inadequate. 
Finding planets on similar orbits around local low-mass stars presents a challenge.  
The lens physical properties are constrained but not directly measured.  
We have discussed additional existing and future data that can measure 
the physical parameters of the lens system directly.

\acknowledgments 
OGLE Team acknowledges Marcin Kubiak and Grzegorz Pietrzy\'nski, former members of the team,
for their contribution to the collection of the OGLE photometric data over the past years. 
Work by RP was partly supported by Polish National Agency for Academic
Exchange grant ``Polish Returns 2019.''
We thank Michael Albrow for information about the \emph{EPOXI} observations.
We thank Ping Chen and Ji Wang for their help. 
This work was supported by NASA contract NNG16PJ32C.  
The OGLE project has received funding from the National Science Centre, Poland,
grant MAESTRO 2014/14/A/ST9/00121 to A.~U. 
The MOA project is supported by JSPS KAKENHI grants No. JSPS24253004, 
JSPS26247023, JSPS23340064, JSPS15H00781, JP16H06287, and JP17H02871. 
The work of N.~K. was supported by JSPS KAKENHI Grant Number JP18J00897.
The work of J.-P.~B. and J.-B.~M. was supported by the ANR COLD-WORLDS project, 
grant ANR-18-CE31-0002-01 of the French Agence Nationale de la Recherche.
Work by C.~H. was supported by the  grants 
of National Research Foundation of Korea (2017R1A4A1015178 and 2019R1A2C2085965). 
This paper includes data gathered with the 6.5-m Magellan Clay Telescope 
at Las Camapanas Observatory, Chile. 
X.~X. and S.~D. acknowledge Project 11573003 supported by NSFC.  This research 
uses data obtained through the Telescope Access Program (TAP), which has been 
funded by 
``the Strategic Priority Research Program-The Emergence of Cosmological Structures'' 
of the Chinese Academy of Sciences (Grant No.11 XDB09000000) and 
the Special Fund for Astronomy from the Ministry of Finance.
Work by A.~G. was supported by AST-1516842 from the US NSF and by JPL grant 1500811.
A.~G. received support from the
European  Research  Council  under  the  European  Union's
Seventh Framework Programme (FP 7) ERC Grant Agreement n. [321035].
Based on data products from observations made with ESO Telescopes at 
the La Silla Paranal Observatory under programme ID~179.B-2002.

\software{
Astropy \citep{astropy13,astropy18},
SExtractor \citep{bertin96},
DoPhot \citep{schechter93}, 
MulensModel \citep{poleski18a_ascl,poleski19a}, 
\url{https://arxiv.org/src/1408.6223v3/anc}, 
MultiNest \citep{feroz08,feroz09}, 
Numerical Recipes \citep{press92fortran}, 
SM \url{https://www.astro.princeton.edu/~rhl/sm/}, 
AstroML \citep{ivezic14},
corner.py \citep{foremanmackey16a}
}

\appendix

\section{  Rejected models }

Tables~\ref{tab:rej}, \ref{tab:cls}, and \ref{tab:source} present parameters of 
rejected wide binary-lens models, close binary-lens models, and a binary-source model, respectively. 
Indexes 1 and 2 indicate paramters for the first and second source, respectively.

\begin{deluxetable*}{lrr}
\tablecaption{Two rejected wide binary-lens models for OGLE-2012-BLG-0838 \label{tab:rej}}
\tablecolumns{3}
\tablewidth{0pt}
\tablehead{
\colhead{Parameter} & \colhead{} & \colhead{}
}
\startdata
$t_0$              & $ 6145.909 \pm 0.029 $  & $ 6145.907 \pm 0.025 $ \\
$u_0$              & $ 0.3585 \pm 0.0017 $   & $ 0.34289 \pm 0.0028 $ \\
$t_{\rm E}$ (d)    & $ 41.53 \pm 0.14 $      & $ 42.88 \pm 0.26 $ \\
$\rho$             & $ 0.00111 \pm 0.00073 $ & $ 0.00372 \pm 0.00020 $ \\
$\alpha$ (deg)     & $ 12.611 \pm 0.046 $    & $ 12.438 \pm 0.044 $ \\
$s$                & $ 2.1199 \pm 0.0044 $   & $ 2.0764 \pm 0.0076 $ \\
$q$                & $ 0.00389 \pm 0.00019 $ & $ 0.00371 \pm 0.00013 $ \\
$F_s/F_{\rm base}$ & $ 0.829 \pm 0.013 $     & $ 0.781 \pm 0.050 $ \\
\hline
$\chi^2/{\rm d.o.f.}$ & $ 835.16 / 678 $ & $ 836.23 / 678 $ \\ 
\enddata
\end{deluxetable*}

\begin{deluxetable*}{lrr}
\tablecaption{Close binary-lens models for OGLE-2012-BLG-0838 \label{tab:cls}}
\tablecolumns{3}
\tablewidth{0pt}
\tablehead{
\colhead{Parameter} & \colhead{close A model} & \colhead{close B model}
}
\startdata
$t_0$              & $ 6146.197 \pm 0.041 $  & $ 6145.866 \pm 0.043 $ \\
$u_0$              & $ 0.3869 \pm 0.0092 $   & $ 0.341 \pm 0.017 $ \\
$t_{\rm E}$ (d)    & $ 39.02 \pm 0.69 $      & $ 44.8 \pm 1.5 $ \\
$\rho$             & $ 0.00123 \pm 0.00012 $ & $ 0.000541 \pm 0.000087 $ \\
$\alpha$ (deg)     & $ 176.68 \pm 0.92 $     & $ 207.20 \pm 0.98 $ \\
$s$                & $ 0.4554 \pm 0.0053 $   & $ 0.4964 \pm 0.0097 $ \\
$q$                & $ 0.0159 \pm 0.0021 $   & $ 0.0120 \pm 0.0016 $ \\
$F_s/F_{\rm base}$ & $ 0.920 \pm 0.030 $     & $ 0.768 \pm 0.049 $ \\ 
\hline
$\chi^2/{\rm d.o.f.}$ &  $ 857.84 / 678 $    & $ 876.34 / 678 $ \\ 
\enddata
\end{deluxetable*}

\begin{deluxetable*}{lr}
\tablecaption{Binary-source model for OGLE-2012-BLG-0838 \label{tab:source}}
\tablecolumns{2}
\tablewidth{0pt}
\tablehead{
\colhead{Parameter} & \colhead{value} 
}
\startdata
$t_{0,1}$ & $ 6145.943 \pm 0.035 $ \\ 
$u_{0,1}$ & $ 0.417 \pm 0.016 $ \\
$t_{\rm E}$ (d) & $ 37.37 \pm 0.95 $ \\
$t_{0,2}$ & $ 6079.3968 \pm 0.0071 $ \\
$u_{0,2}$ & $ 0.00037 \pm 0.00028 $ \\
$\rho_2$ & $ 0.00781 \pm 0.00029 $ \\
$F_{s,1}/F_{\rm base}$ & $ 1.027 \pm 0.052 $ \\
$F_{s,2}/F_{\rm base}$ & $ 0.00875 \pm 0.00026 $ \\
\hline
$\chi^2/{\rm d.o.f.}$ &  $ 988.98 / 678 $ 
\enddata
\end{deluxetable*}

\bibliography{paper}

\end{document}